\documentclass[showpacs,prb,amsmath]{revtex4}
\usepackage{graphicx}
\usepackage{bm}
\begin{document}
\preprint{}
\title{The Role of Non-Equilibrium Carriers in Formation of Thermo-E.M.F. in
Bipolar Semiconductors}
\author{Yu.~G.~Gurevich}
\email{gurevich@fis.cinvestav.mx}
\author{G.~N.~Logvinov}
\email{logvinov@fis.cinvestav.mx}
\author{I.~N.~Volovichev}
\email{vin@ire.kharkov.ua}
\altaffiliation{Permanent address: Institute for Radiophysics and 
Electronics, Academy of Sciences of Ukraine, Kharkov 310085, Kharkov 
310085, Ukraine.}
\affiliation{Departamento de F\'{\i}sica, CINVESTAV---IPN,\\
Apdo. Postal 14--740, 07000 M\'{e}xico, D.F.}
\author{G.~Espejo}
\email{gespejo@fis.cinvestav.mx}
\affiliation{CICATA---IPN, Garc\'{\i}a Obeso 306, Col.~Centro,\\
58000, Mor., Mich., M\'exico.}
\author{O.~Yu.~Titov}
\email{oleg.titov@aleph-tec.com}
\affiliation{CICATA---IPN, Jos\'{e} Siurob 10, Col.~Alameda,\\
76040 Santiago de Quer\'{e}taro, Qro., M\'{e}xico.}
\author{A.~Meriuts}
\email{meriuts@kpi.kharkov.ua}
\affiliation{Kharkov Politechnical University, Kharkov 310002, Ukraine.}
\date{\today}

\begin{abstract} 
It is presented a  new approach to thermoelectric phenomena, as a linear 
transport process of non-equilibrium charge carriers. The role of 
non-equilibrium carriers, as well as surface and bulk recombination, has 
shown to be crucial even within a linear approximation. Electron and hole 
Fermi quasi-levels that appeared in a thermal field are calculated for 
the case of thermoelectric current flow through a circuit and the 
corresponding boundary conditions are obtained. It is shown for the first 
time, that the Fermi quasi-level of one of the subsystems of 
quasi-particles, can be a non-monotonic function of the coordinates. 
General expressions for the thermoelectric current, thermo-e.m.f., and 
electrical resistance of bipolar semiconductors have been obtained. For 
the first time, surface recombination and surface resistance were taken 
into account in thermoelectric phenomena.
\end{abstract} 


\maketitle

\section{Introduction} 
The critical treatment of the generally accepted theory of 
thermo-e.m.f. in bipolar semiconductors was presented in theoretical
works by Gurevich {\em et el.}\cite{Lyubimov,Gurevich} The calculations
were carried out without recombination term (the scope of those papers
was intentionaly limited to explain the correct definition of
thermo-e.m.f.).

We will continue our constructive criticism (as we will offer some
practical solutions) with following, never-discussed facts.

The first of them is related to applied temperature difference. As soon
as we have the temperature gradient, both electrons and holes diffuse
from the hot end of the sample to the cold one. As a result, electron
and hole concentrations should increase on the cold end of the sample if
recombination processes are ignored (usually they even were not
discussed in the study of thermoelectric phenomena,\cite{Anselm} and
corresponding terms were not presented in the equations. Therefore one
can say that recombination was ignored.). 
 
As a consequence, the well known correlation\cite{Anselm} 
\begin{equation} 
n_{0} p_{0} = n_{i}^2 
\label{ec:conc} 
\end{equation} 
is violated, where $n_{0}$, $p_{0}$ are equilibrium electron and hole 
concentrations at some equilibrium temperature, $n_{i}$ is the intrinsic 
concentration in the same semiconductor. 
 
The fact that Eq.~({\ref{ec:conc}}) be correct, is based on the existence of 
common Fermi levels for electrons and holes and the equality 
\begin{equation} 
\mu_{n}^0 = -\varepsilon_{g} - \mu_{p}^0, 
\label{ec:chempot} 
\end{equation} 
always suits in this case. Here $\mu_{n}^0$, $\mu_{p}^0$ are electron
and hole chemical potentials in equilibrium state, $\varepsilon_{g}$ is
the energy gap. 
 
So if equation ({\ref{ec:conc}}) is not fulfilled, then $\mu_{n} \neq
- \varepsilon_{g} - \mu_{p}$, where $\mu_{n}$ and $\mu_{p}$ are the
non-equilibrium electron and hole chemical potentials. This means that 
the common Fermi level is splited and two Fermi quasi-levels (electron
and hole) originate. 

Since the internal thermoelectric field is determined by the gradient of
electrochemical potential,\cite{Logvinov} the question that immediately
arose was: Which field shall we use to calculate thermo-e.m.f.: $E_{1} =
-\nabla(\varphi - \mu_{n}/e)$ or $E_{2} = -\nabla(\varphi +
\mu_{p}/e)$? Here, $\varphi$ is the electrical potential caused by the
redistribution of electric charges in the thermal field, and $-e$ is the
electron charge. Besides, recombination
processes shall be taken into account  if there appear non-equilibrium carriers.
  
The second aspect is related to the way that thermo-e.m.f. is usually
calculated in an open circuit. At the same time, it is well known that the
correct definition of e.m.f. of any nature shall be related to a closed
circuit. This circumstance is especially important to determine
thermo-e.m.f. in bipolar semiconductors. It is enough to imagine the
particular special situation with inhomogeneously heated $p$-type
semiconductor closed with the metal section of circuit. There are no holes
in the metal and there are no electrons in a $p$-type semiconductor. The
explanation of this phenomena can not be defined within the limits of
thermoelectric traditional theory. It is clear that it is necessary to use
boundary conditions taking into account current flux and surface
recombination processes. Nobody had done that before.
 
The third aspect is that the consideration of thermo-e.m.f. in bipolar
semiconductors is carried out, usually, after the study of this problem in
unipolar semiconductors, and many results are obtained ``by analogy.''
Methodically it ought to be done starting from ``point zero'' to avoid a
series of mistakes and incorrect judgments.
 
There are more aspects related with this problem but it is clear that one
of the main problems when establishing a correct theory of thermo-e.m.f.
in bipolar semiconductors is the successive accounting of the
non-equilibrium carriers in electric and thermal currents caused by
inhomogeneous heating of the semiconductor.
 
This problem has been partially studied;\cite{Lyubimov,Gurevich} however,
new ideas have arisen: the modernization in description of recombination
processes, the accurate definition of current boundary conditions,
calculation of spatial dependencies of Fermi quasi-levels, consideration
of general expressions for thermo-e.m.f., etc...

The need of more precise theories and the generalization of some results,
considered earlier, have just appeared.
 
Covering all aforementioned problems is beyond the scope of present work.
We will only expatiate on those of common interest.
 
\section{The Main Equations of the Problem} 
\begin{figure}[!ht]
\centering
\includegraphics[width=8cm, height=5cm]{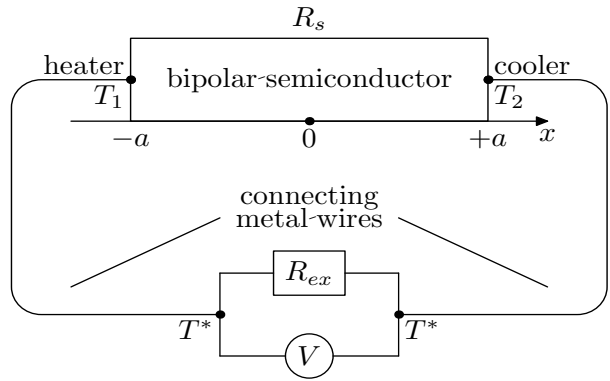}
\caption{Electrical circuit for the measurement of thermo-e.m.f., $R_{s}$
is the semiconductor resistance.}
\label{f:fig1}
\end{figure}

Let us consider the model of thermoelectric circuit, consisting of an 
isotropic, homogeneous bipolar semiconductor, of parallelepiped form,
closed by a metal section with an electrical conductivity $\sigma_{m}$, 
length $L$ (see Fig.~\ref{f:fig1}) and unit cross-section. 
The resistance of external circuit (including voltmeter's resistance) is 
$R_{ex}= L/\sigma_{m}$. The metal-semiconductor contact on the left, 
$x=-a$, is kept under a temperature $T_{1}$; at $x=+a$ is at a temperature 
$T_{2} < T_{1}$. Here and below,  temperature is expressed in energy 
units, {\it i.~e.} Boltzmann's constant is equal to unit. The
lateral sides are thermally insulated, so the problem is unidimensional. 
We suppose that the temperature difference $\Delta T = 
T_{1} - T_{2}$ is small enough [$\Delta T/T^{*} \ll 1, T^{*}= 
(T_{1}+T_{2})/2$], so the problem is linear. The thermal contacts
between metal and semiconductor are assumed to be isothermal, for
simplicity. We also assume that non-equilibrium temperatures of all 
quasi-particle subsystems (electrons, holes and phonons) taking part in
heat transport are the same. 
 
Under these conditions, the temperature field in the semiconductor is 
represented by the function 
\begin{equation} 
T(x) = T^{*} -\frac{\Delta T}{2a} x . 
\label{ec:temp} 
\end{equation} 
 
At stationary state, the thermo-e.m.f. is generated in the circuit
and the constant thermoelectric current $j_{0}$ flows through it. In
bipolar semiconductors, 
\begin{equation} 
j_{0} = j_{n} + j_{p} , 
\label{ec:fcurrent} 
\end{equation} 
where 
\begin{equation} 
\begin{gathered} 
j_{n} = 
 -\sigma_{n}\left(\frac{d\tilde\varphi_{n}}{dx} + 
\alpha_{n}\frac{dT}{dx}\right), \\ 
j_{p} = 
 -\sigma_{p}\left(\frac{d\tilde\varphi_{p}}{dx} + 
\alpha_{p}\frac{dT}{dx}\right) 
\end{gathered} 
\label{ec:parjnp} 
\end{equation} 
are the partial electron and hole currents; $\sigma_{n}$, $\sigma_{p}$
are the bulk electron and hole electrical conductivities; $\alpha_{n}$,
$\alpha_{p}$ are the electron and hole thermoelectric powers;
$\tilde\varphi_{n,p} = \varphi \mp \mu_{n,p}/e$ are the electrochemical 
electron and hole potentials (Fermi quasi-levels). 
 
The equations of continuity for partial currents $j_{n}$ and $j_{p}$
are:
\begin{equation} 
\frac{d j_{n}}{dx} = eR_{n}, \qquad \frac{d j_{p}}{dx} = -eR_{p} , 
\label{ec:pcurrent} 
\end{equation} 
where $R_{n,p}$ are bulk recombination rates. 

In accordance with Eq.~ $\frac{\partial\rho }{\partial t}  = - 
\operatorname{div} \bm{j} $ (the consequence of Maxwell equation) in 
the static case we obtain,
\begin{equation} 
\operatorname{div} \bm{j} = 
\operatorname{div} \left(\bm{j}_{n} + \bm{j}_{p}\right) = 0 . 
\label{ec:divj} 
\end{equation} 
where $\rho$ is charge density, $\bm{j}$ is the total current.

Let us note that  Eqs.~(\ref{ec:pcurrent}) and \ref{ec:divj}) are 
independent, so the electron and hole bulk recombination rates ($R_{n}$, 
$R_{p}$) must be the same.\cite{VolGur}
 
By itself, the correct description of the recombination term $R_{n,p}$ in
Eq.~(\ref{ec:pcurrent}) is a serious problem in the presence of
temperature fields. The frequently used form for $R$: $R_n ={\delta
n}/{\tau_n}$, $R_p={\delta p}/{\tau_p}$ does not ensure the identical
equality of the recombination rates of electrons and holes for any
recombination mechanism. In its turn, non-equal recombination rates do not
preserve the total charge [see Eq.~(\ref{ec:divj})]. Moreover, treating
the equality $R_{n}=R_{p}={\delta p}/{\tau_{p}}$ as an additional
condition to find the non-equilibrium concentration, makes no physical
sense.\cite{VolGur,Volovichev}
 
It is necessary to use the expression obtained from statistical
consideration of electron transitions between valence and conduction bands
(or the Shockley-Read expression for recombination through impurity
levels) to obtain a correct description of
recombination.\cite{VolGur,Volovichev}
 
For the first case, for example, $R_n \equiv R_p = k(np - n_i^2)$ (where 
$k$ is the capture factor, $n=n_{0} + \delta n'$, $p=p_{0}+ \delta p'$),
$\delta n'$, $\delta p'$ are non-equilibrium additions to the
equilibrium concentrations $n_{0}$ and $p_{0}$. 
 
Assuming $\delta n' \ll n_{0}$, $\delta p' \ll p_{0}$ and linearizing
the recombination rate expression, we get 
\begin{equation} 
R = R_{n} = R_{p} = 
 \frac{\delta n'}{\tau_n} + \frac{\delta p'}{\tau_p} , 
\label{ec:recom} 
\end{equation} 
where $\tau_n = (k p_{0})^{-1}$, $\tau_p = (k n_{0})^{-1}$. 
 
The values $\tau_n$ and $\tau_p$ have dimensions of time but have not some
physical sense for the general case ($\delta n' \neq \delta p'$). Only if
the condition of quasi-neutrality is fulfilled,\cite{Silin} ($\delta n' =
\delta p'$), the uniform time $\tau^{-1} = \tau_{n}^{-1}+\tau_{p}^{-1}$
can be introduced, which has the sense of life-time of non-equilibrium
carriers.
 
Equation~(\ref{ec:pcurrent}) must be supplemented by the necessary 
boundary conditions. Since holes can not pass trough semiconductor-metal
contacts, boundary conditions for them are: 
\begin{equation} 
\left. j_{p}\right|_{x=\mp a} = \mp eR_{s}, 
\label{ec:front1} 
\end{equation} 
where $R_{s}$ is the surface recombination rate. 
 
The boundary conditions for the electron current $j_{n}$, taking into 
account Eq.~(\ref{ec:fcurrent}), are in the form 
\begin{equation} 
\left. j_{n}\right|_{x=\mp a} = j_{0} \pm eR_{s} . 
\end{equation} 
 
Similarly to the bulk recombination, the expressions for surface 
recombination must be written in the following form:
\begin{equation} 
R_{n}^s = R_{p}^s = R_{s} = S_{n} \delta n' + S_{p}\delta p' , 
\end{equation} 
where $S_{n}$ and $S_{p}$ are some coefficients which characterize the
properties of semiconductor surface. Only when the quasi-neutrality
condition takes place ($\delta n' = \delta p'$), the value $S=S_{n} +
S_{p}$ acquires the sense of surface recombination velocity, and 
\begin{equation} 
R_{s} = S\delta n' . 
\end{equation} 
 
It should be noted that the full current $j_{0}$ satisfies the 
following condition at the metal-semiconductor contacts:\cite{Gurevich} 
\begin{equation} 
j_{0} = \pm \sigma_{n}^{s_{\pm}} 
 \left[ 
\varphi_{s}(\pm a) - \varphi_{m}(\pm a) 
 \right] 
\mp 
\frac{\sigma_{n}^{s}}{e} 
 \left[ 
\mu_{n}^{s}(\pm a) - \mu_{m} 
 \right] 
\pm \frac{\sigma_{n}^{s}}{e}\Delta\varepsilon_{c}
.
\label{ec:front2} 
\end{equation} 
 
Here, $\sigma_{n}^{s}$ is the surface electrical conductivity; 
$\varphi_{s}(\pm a)$, $\varphi_{m}(\pm a)$ are the electrical potentials
for semiconductor and metal surfaces, respectively; $\mu_{n}^{s}(\pm a)$
is the surface chemical potential of electronsThe ; $\mu_{m}$ is the
chemical potential of metal; $\Delta\varepsilon_{c}$ is the energy gap
between the bottom of the conducting band of the semiconductor and the
metal conduction band, at the semiconductor-metal contact. 
 
Finally, the electrical potential $\varphi_{s}(x)$ is determined from 
Poisson equation if quasi-neutrality is absent:
\begin{equation} 
\frac{d^2\varphi_{s}(x)}{dx^2} = -4\pi\rho , 
\label{ec:poisson} 
\end{equation} 
where $\rho(x)= -e\left[\delta n'(x)-\delta p'(x)\right]$ is the bulk 
charge density. 

The continuity of electrical and electrochemical potentials at the $x=\pm a$
should serve as a boundary condition for equation (\ref{ec:poisson}):
\begin{eqnarray} 
\begin{gathered} 
\varphi_{s}(\mp a)=\varphi_{m}(\mp a) \\ 
\label{ec:contpota} 
\tilde\varphi_{s}(\mp a)=\tilde\varphi_{m}(\mp a) 
\label{ec:contpotb} 
\end{gathered} 
\end{eqnarray} 

These potentials are discussed in the next section. 

\section{Inhomogeneous Thermodynamic Equilibrium Distributions of
Electrical and Chemical Potentials}
\begin{figure}[!ht]
\centering
\includegraphics[width=8cm, height=5cm]{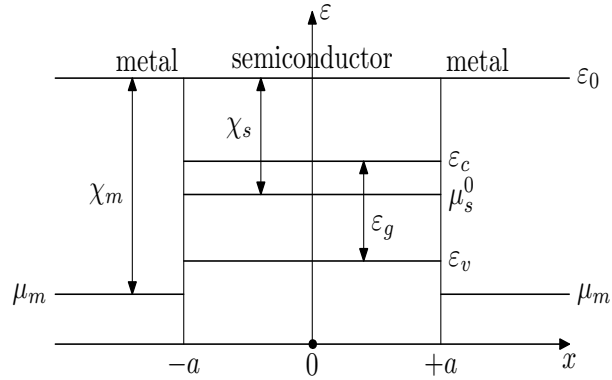}
\caption{Energy diagram before joining semiconductor and metal in
circuit. $\varepsilon_{0}$ is the vacuum level; $\chi_{s}$, $\chi_{m}$
are semiconductor and metal work functions; $\varepsilon_{c}$,
$\varepsilon_{v}$ are the bottom of conduction band and the top of
valence band in semiconductor, respectively; $\mu_{s}^0$, $\mu_{m}$ are the
chemical potentials for semiconductor and for metal, $\varepsilon_{g}$ is
the band gap.}
\label{f:fig2}
\end{figure}

First of all, let us consider the energy diagram of thermoelectric 
circuit elements which are not joined yet into a single whole (see 
Fig.~\ref{f:fig2}). 

Electron and hole concentrations in bipolar semiconductors are 
represented by the following well known expressions:\cite{Anselm} 
\begin{equation} 
\begin{gathered} 
n_{0}(T^{*}) = 
 \gamma_{n}(T^{*})\exp\left[\frac{\mu_{n}^0(T^{*})}{T^{*}}\right] \\ 
p_{0}(T^{*}) = 
 \gamma_{p}(T^{*})\exp\left[\frac{\mu_{p}^0(T^{*})}{T^{*}}\right] . 
\end{gathered} 
\end{equation} 
 
Here, $\mu_{n}^0(T^{*})=\mu_{s}^0$; the hole chemical potential is
related to the electron chemical potential $\mu_{n}^0$ by the expression 
(\ref{ec:chempot}); $\gamma_{n,p}(T^{*})= 
\frac{1}{4}\left(\frac{2m_{n,p}T^{*}}{\pi \hbar^{2}} \right)^{3/2}$ 
are the electron and hole density of states at the bottom of the
conduction band and at the top of the valence band; $m_{n,p}$ are the
electron and hole effective masses. 
 
A new thermodynamic equilibrium state arises after the creation of a circuit
like that shown in Fig.~\ref{f:fig1}, but in the absence of temperature
difference $\Delta T$. This state is characterized by a common
temperature $T^{*}$, and a common electrochemical potential
$\tilde\varphi(x)= \varphi(x)- \mu_{s}^0/e$, where $\varphi(x)$ is the
electrical potential that arises as a result of charge redistribution
between semiconductor and metal, due to the difference in work function. 
 
Usually, one can disregard this charge redistribution in the metal, as it
takes place within some atomic layers and one may consider that whole
charge is concentrated at the metal surface. The semiconductor charge
redistribution essentially depends on the correlation between
semiconductor thickness and the Debye radius $r_{d}^2 = T^{*}/4\pi e^2
n_{0}(T^{*})$ (for simplicity, we omit the dielectric constant). 
 
\begin{figure}[!ht]
\centering
\includegraphics[width=8cm, height=5cm]{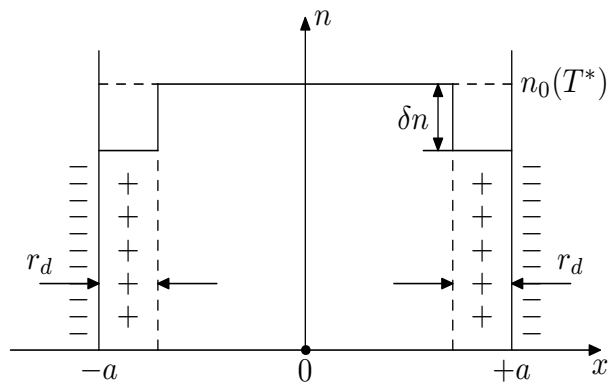}
\caption{Distribution of electron concentration in semiconductor
near the contact ends with metal $x = \mp a$, at $\chi_{s} <
\chi_{m}$. Typical value of electron concentration $n_{0} = 10^{16} 
cm^{-3}$ an$= 10^{12} cm^{-3}$. Signs (+) and (-) point the charge 
distributions in the quasi-neutral case ($r_{d}\rightarrow 0$)} 
\label{f:fig3}
\end{figure}

The condition of quasi-neutrality is fulfilled if $a^2 \gg r_{d}^2$. It
is characterized by the charge redistribution only in the layers close
to the semiconductor surface. The thickness of these layers is about 
$r_{d}$ and for typical semiconductors is approximately equal
to $10^{-5}$--$10^{-7}$cm. The change of charge concentration $\delta
n_{0}(T^{*})$ in these layers is small in comparison with the
equilibrium concentration. At the same time, this redistribution
creates a highly appreciable electric field. The concentration
distribution in the semiconductor is represented schematically in
Fig.~\ref{f:fig3}. 
 
\begin{figure}[!ht]
\centering
\includegraphics[width=8cm, height=5cm]{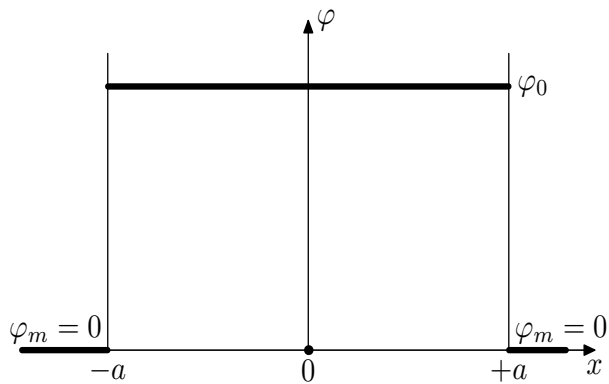}
\caption{Contact voltage in the quasi-neutral approximation.}
\label{f:fig4}
\end{figure}

In the limiting case of a long sample ($r_{d}\rightarrow 0$), as it is
implied in the quasi-neutrality model, the double electrical layer is
generated, and electrical potential undergoes a gap, which is shown in
Fig.~\ref{f:fig4}. The contact voltage arises. 

\begin{figure}[!ht]
\centering
\includegraphics[width=8cm, height=5cm]{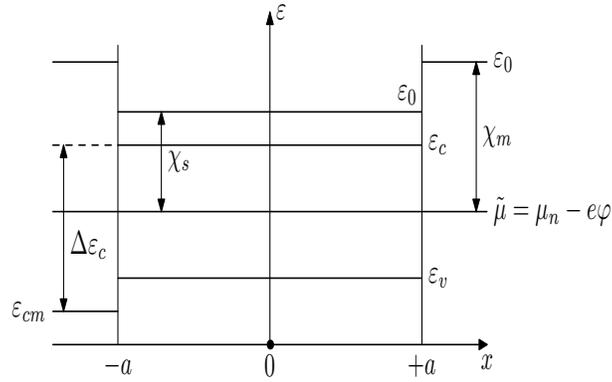}
\caption{Energy diagram once equilibrium state has been 
reached. $\varepsilon_{cm}$ is the bottom of metal conduction band,
$\Delta\varepsilon_{c}$ is the energy gap between the bottom 
of semiconductor conduction band and metal conduction band.}
\label{f:fig5}
\end{figure}

The energy diagram of the circuit in this case is represented in
Fig.~\ref{f:fig5}. 

The solution of Poisson's equation is not required under quasi-neutrality 
approximation and the contact voltage $\varphi_{0}$ (the potential of
metal is equal to zero), can be found from the condition of equality of 
electrochemical potential on the semiconductor-metal contact (see 
Fig.~\ref{f:fig5}): 
\begin{equation} 
\varphi_{0} = 
\frac{1}{e}\left[\mu_{n}^0(T^{*})- \mu_{m} - \Delta\varepsilon_{c}\right] 
\end{equation} 

The appearance of term $\Delta\varepsilon_{c}$ is explained by the 
need of choosing a common reference energy level. We took as
reference point, the bottom of the conduction band. 

\begin{figure}[!ht]
\centering
\includegraphics[width=8cm, height=5cm]{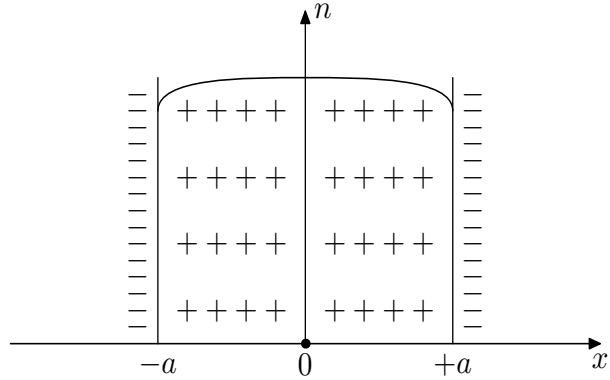}
\caption{Distribution of charge in the absence of quasi-neutrality.}
\label{f:fig6}
\end{figure}

A completly different case can be watched in semiconductors which thicknesses
are about (or less than) the Debye radius ($a^2\lesssim
r_{d}^2$), i.e. when the quasi-neutrality condition is absent. Now, the
charge redistribution takes place in the whole semiconductor's bulk and
the double charge layer does not appear (see Fig.~\ref{f:fig6}). 

Electrical potential is continuous at every point including the points
$x=\mp a$. These continuities form boundary conditions 
(\ref{ec:contpotb}) to Poisson's equation. 

Let us write the distributions of chemical potentials, electron and hole
concentrations in the following form: 
\begin{equation} 
\begin{gathered} 
\mu_{n}(x) = \mu_{n}^0(T^{*}) + \delta\mu_{n}'(x) \\ 
\mu_{p}(x) = \mu_{p}^0(T^{*}) + \delta\mu_{p}'(x) \\ 
n(x) = n_{0}(T^{*}) + \delta n_{0}'(x) \\ 
p(x) = p_{0}(T^{*}) + \delta p_{0}'(x) , 
\end{gathered} 
\end{equation} 
where $\delta\mu_{n}'(x)$, $\delta\mu_{p}'(x)$, $\delta n_{0}'(x)$, 
$\delta p_{0}'(x)$ are unknown functions.

We can consider that $\delta\mu_{n}'(T^{*})/T^{*} \ll 1$, if the
difference between work functions is a small one
$(\chi_{m}-\chi_{s})/\chi_{m}\ll 1$, or if the temperature $T^{*}$ is a
high one.

Then, from 
\begin{equation*} 
n(x) 
= 
\gamma_{n} 
\exp\left[\frac{\mu_{n}^0(T^{*})}{T^{*}}\right] 
\exp\left[\frac{\delta\mu_{n}'(x)}{T^{*}} \right]
= n_{0}\left[1 + \frac{\delta\mu_{n}'(x)}{T^{*}}\right] 
,
\end{equation*} 
it follows that 
\begin{equation} 
\delta n_{0}'(x) = \frac{n_{0}}{T^{*}}\delta\mu_{n}'(x) .
\label{ec:dn} 
\end{equation} 

By analogy,
\begin{equation} 
\delta p_{0}'(x) = \frac{p_{0}}{T^{*}}\delta\mu_{p}'(x) . 
\label{ec:dp}
\end{equation} 

We examine the equilibrium state [see Eq.~(\ref{ec:chempot})], thus 
\begin{equation} 
\delta\mu_{n}'(x) = - \delta\mu_{p}'(x) 
\end{equation} 
and 
\begin{equation} 
\delta p_{0}'(x) = - \frac{p_{0}}{T^{*}}\delta\mu_{n}'(x) . 
\end{equation} 

In this case, Poisson's equation is reduced to 
\begin{equation} 
\frac{d^2\varphi (x)}{dx^2} = \frac{1}{er_{d}^2}\delta\mu_{n}'(x) , 
\label{ec:pois1} 
\end{equation} 

Here, $r_{d}^2 = T^{*}/4\pi e^2(n_{0}+p_{0})$, $\varphi (x)$ is the
potential distribution in the semiconductor. 

The electric current absence condition leads to a correlation 
between electrical and chemical potentials: 
\begin{equation} 
\frac{d\varphi(x)}{dx} = \frac{1}{e}\frac{d\mu_{n}'(x)}{dx} . 
\label{ec:pois2} 
\end{equation} 

Solving Eqs.~(\ref{ec:pois1},\ref{ec:pois2}) simultaneously, we obtain: 
\begin{eqnarray} 
\begin{gathered} 
\delta\mu_{n}'(x) = C_{1}e^{x/r_{d}} + C_{2}e^{-x/r_{d}} \\ 
\varphi_{0}(x) = \frac{1}{e}\left( C_{1}e^{x/r_{d}} + C_{2}e^{-x/r_{d}} 
\right) + C_{3} 
\end{gathered} 
\end{eqnarray} 
where $C_{1,2,3}$ are yet to be fixed. 

To determine them, the constancy of the electrochemical 
potentials and the continuity of the electrical potential at the boundaries 
$x=\pm a$ may be used: 
\begin{equation} 
\begin{gathered} 
\pm \varphi (\pm a) 
\mp \frac{1}{e}\left[\mu_{n}^{0}(T^{*}) + 
\delta\mu_{n}'(\pm a) - 
\mu_{m} \right]
\pm \frac{\Delta\varepsilon_{c}}{e} = 0 , \\
\varphi (\pm a) = 0
.
\end{gathered} 
\end{equation} 

After taking into account the values for the previously undefined 
constants, we get: 
\begin{eqnarray} 
\begin{gathered} 
\varphi (x) = 
\varphi_{0}\left[1 - \frac{ch (x/r_{d})}{ch (a/r_{d})} \right] , 
\label{ec:potconta}\\ 
\delta\mu_{n}'(x) = - e \varphi_{0}\frac{ch (x/r_{d})}{ch (a/r_{d})} . 
\end{gathered} 
\label{ec:potcontb} 
\end{eqnarray} 

\begin{figure}[!ht]
\centering
\includegraphics[width=8cm, height=5cm]{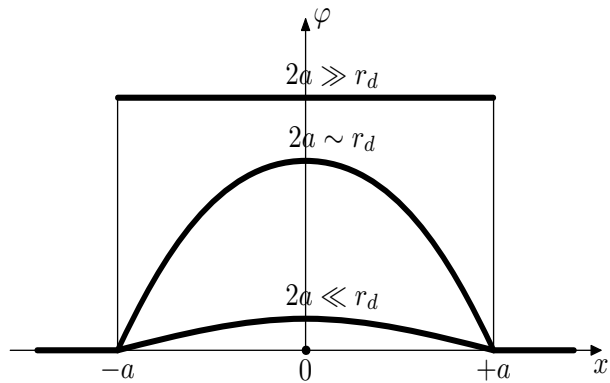}
\caption{Distribution of the electrical potential for different
semiconductor lengths.}
\label{f:fig7}
\end{figure}

The distribution $\varphi (x)$ [see Eq.~(\ref{ec:potconta})] is shown in
Fig.~\ref{f:fig7}. 

It can be seen that the electrical potential distribution is quite
different from the distribution when quasi-neutrality is present. The
conception of contact voltage lacks of meaning in general, and exists
only at $r_{d} \ll a$. 

\begin{figure}[!ht]
\centering
\includegraphics[width=8cm, height=5cm]{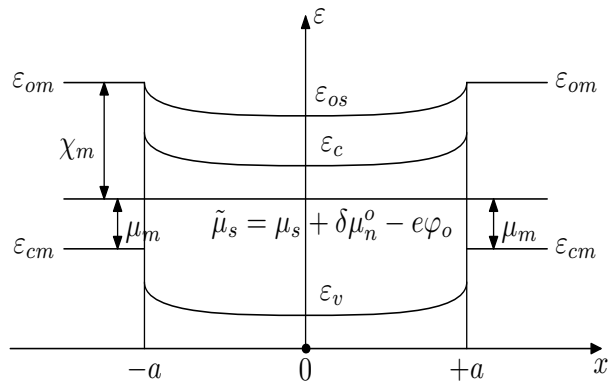}
\caption{Energy diagram in the absence of the 
quasi-neutrality condition. $\varepsilon_{om}$ is the vacuum level for 
both metal and semiconductor.}
\label{f:fig8}
\end{figure}

The energy diagram of the circuit is represented schematically in 
Fig.~\ref{f:fig8}. 

\section{Fermi Quasi-levels and General Expressions for Thermo-E.M.F. 
under Quasi-neutrality Condition}
From the previous section, it is clear that electrical potential 
distribution and chemical potential are strongly different if 
quasi-neutrality is present or absent (even for the case $\Delta T =0$). 
Therefore, later on this paper we will restrict to quasi-neutrality case 
only. The opposite situation, when quasi-neutrality is absent, is treated 
somewhere else. 

It is important to understand the double role of temperature 
inhomogeneity in the construction of a thermo-e.m.f. theory. On the one
hand, it is the cause of inhomogeneity in electrical, chemical
potentials, and concentration in homogeneous semiconductors. On the 
other hand, temperature, namely its gradient, determines the 
thermodynamic force causing particle motion within the sample [terms
$\alpha_{n,p}\nabla T$ in Eqs.~(\ref{ec:parjnp})]. This understanding
allows to imagine the process of forming thermo-e.m.f. and
thermoelectric current in several stages. 

At the first stage, the temperature is the same and equal to $T^{*}$ at 
all points of the thermoelectric circuit. Only contact voltage appears
in the vicinity of surfaces $x=\mp a$. This situation was described in
the previous section in detail. 

At the second stage, we suppose that temperature difference $\Delta T$ is
created between planes $x=\mp a$ and temperature distribution is
given by the function (\ref{ec:temp}). The metal chemical potential does
not depend on temperature and remains unchanged, the chemical potential
and the carrier concentration of semiconductor are the only ones that
changes. The initial inhomogeneous electron and hole potentials, and their
concentrations, are formed at this stage: 
\begin{equation} 
\begin{gathered} 
\mu_{n,p}(x) = \mu_{n,p}^0(T^{*}) + \delta\mu_{n,p}^0(x) \\ 
n_{0}(x) = n_{0}(T^{*}) + \delta n_{0}(x), \\ 
p_{0}(x) = p_{0}(T^{*}) + \delta p_{0}(x) 
\end{gathered} 
\end{equation} 
where $\delta\mu_{n}^0(x)$, $\delta\mu_{p}^0(x)$, $\delta n_{0}(x)$, 
$\delta p_{0}(x)$ are inhomogeneous additions to equilibrium chemical 
potentials and equilibrium concentrations. These additions arise due to 
 inhomogeneous temperature [see Eq.~(\ref{ec:temp})]. In spite of this,
we suggest that $\nabla T$, as a force, has not ``switched on'' yet and
the carriers are immovable. 

At the third stage $\nabla T$ is in the condition of ``switched off'' too,
but electrons and holes are free and redistributed due to diffusion 
flux, and built-in thermoelectric field appears. The electrochemical 
potentials become common and constant and ``equilibrium'' state is
reached. Inverted comas are used here because it is impossible to use
the term  ``equilibrium'' in presence of a temperature field
(\ref{ec:temp}). This situation is similar to the formation of
inhomogeneous chemical potential and concentration in the
inhomogeneously doped semiconductor: at some point, inhomogeneously
distributed donors (for example), give electrons to conduction band
and form the initial electron distribution. In the next moment, this
distribution changes due to arising diffusion current and internal
electric field. As a result, the equilibrium state of inhomogeneous
distributed charge carriers is established. The main difference of this
example from a distribution that has been observed in the temperature
field, is that this is the true equilibrium distribution. 

The processes included in the second and third stages were described in
Ref.~\onlinecite{Gurevich}, so we do not consider them here in detail
and we only use some results.  Let us note that at this stage,
concentration only determines equilibrium concentration which is used under
recombination process description. 

At the last stage, the terms $\alpha_{n,p}\frac{dT}{dx}$ are ``switched
on,'' the new additional chemical, electrical potentials, and 
concentration distributions appear and the thermoelectric current flows 
through the circuit. 

Let us present the concentrations and chemical potentials in this case
as follows: 
\begin{equation} 
\begin{gathered} 
n(x) = n_{0}(T^{*}) + \delta n_{0}(x) + \delta n(x) \\ 
p(x) = p_{0}(T^{*}) + \delta p_{0}(x) + \delta p(x), \\ 
\mu_{n}(x) = \mu_{n}^0(T^{*}) + \delta\mu_{n}^0(x) + \delta\mu_{n}(x)\\ 
\mu_{p}(x) = \mu_{p}^0(T^{*}) + \delta\mu_{p}^0(x) + \delta\mu_{p}(x) 
\end{gathered} 
\end{equation} 
where\cite{Gurevich} 
\begin{align}
\delta n_{0}(x) & = 
n_{0}(T^{*})
 \left\{ 
\left[ 
\frac{\mu_{n}^{0}(T^{*})}{T^{*}} - \frac{3}{2} 
\right] 
\frac{\Delta T}{T^{*}} \frac{x}{2a} + 
\frac{\delta\mu_{n}^{0}(x)}{T^{*}} 
\right\}\nonumber\\ 
\delta p_{0}(x) & = p_{0}(T^{*}) 
\left\{ 
\left[ 
\frac{\mu_{p}^{0}(T^{*})}{T^{*}} - \frac{3}{2} 
\right] 
\frac{\Delta T}{T^{*}} \frac{x}{2a} 
 +
\frac{\delta\mu_{p}^{0}(x)}{T^{*}} 
\right\}\\ 
\delta\mu_{n}^{0}(x) & = -\frac{d\mu_{n}^{0}(T^{*})}{dT^{*}}\frac{\Delta T}{2a} x \nonumber\\ 
\delta\mu_{p}^{0}(x) & = - \delta\mu_{n}^{0}(x) .\nonumber 
\end{align} 

The spatial charge redistribution in turn leads to the electric 
potential change: 
\begin{equation} 
\varphi(x) = 
\varphi_{0} + \delta\varphi_{0} + \delta\varphi_{1}(x) + \delta\varphi(x). 
\end{equation} 

Here\cite{Gurevich} 
\begin{equation} 
\begin{gathered} 
\delta\varphi_{0} = \frac{1}{e}\delta\mu_{n}^{0}(-a)\\ 
\delta\varphi_{1} = - \frac{1}{e} \left[ \delta\mu_{n}^{0}(-a) - \delta\mu_{n}^{0}(x) \right] 
. 
\end{gathered} 
\end{equation} 

If the condition of quasi-neutrality is imposed [$\delta n(x) = \delta 
p(x)$] then it is easy to get that 
\begin{equation} 
\delta\mu_{p}(x) = \frac{n_{0}(T^{*})}{p_{0}(T^{*})}\delta\mu_{n}(x) 
. 
\end{equation} 

Now, the expressions for electron and hole currents are: 
\begin{equation} 
\begin{gathered}
j_{n}(x) = \sigma_{n} \left\{- \frac{d}{dx} \left[ \delta\varphi(x) - 
\frac{1}{e}\delta\mu_{n}(x) \right] - \alpha_{n} \frac{dT}{dx} \right\} \\ 
j_{p}(x) = \sigma_{p} \left\{- \frac{d}{dx} \left[ \delta\varphi(x) + 
\frac{n_{0}(T^{*})}{p_{0}(T^{*})}\frac{\delta\mu_{n}(x)}{e} \right] - \alpha_{p} \frac{dT}{dx} \right\}
. 
\end{gathered} 
\end{equation}

Substituting these expressions into Eqs.~(\ref{ec:pcurrent}) we get the 
following system of equations: 
\begin{align} 
\frac{d^2\delta\varphi(x)}{dx^2}
 - \frac{1}{e} \frac{d^2\delta\mu_{n}(x)}{dx^2} 
&=
 -e \frac{n_{0}(T^{*})}{T^{*}} \frac{\delta\mu_{n}(x)}{\tau\sigma_{n}} \label{ec:ec}\\ 
\frac{d^2\delta\varphi(x)}{dx^2}
 + \frac{1}{e}\frac{n_{0}(T^{*})}{p_{0}(T^{*})} \frac{d^2\delta\mu_{n}(x)}{dx^2}
&=
 +e \frac{n_{0}(T^{*})}{T^{*}} \frac{\delta\mu_{n}(x)}{\tau\sigma_{p}}, \nonumber 
\end{align} 
where 
\begin{equation} 
\tau= \frac{\tau_{n}\tau_{p}}{\tau_{n} + \tau_{p}} 
, 
\end{equation} 
and times $\tau_{n}$, $\tau_{p}$ were determined in Eq.~(\ref{ec:recom}). 

The solution of system (\ref{ec:ec}) is: 
\begin{equation} 
\begin{gathered} 
\delta\mu_{n}(x) = C_{1}'e^{\lambda x} + C_{2}'e^{-\lambda x} \\ 
\delta\varphi(x) = \frac{\beta^2}{e\lambda^2}\delta\mu_{n}(x) + C_{3}'x + C_{4}' ,
\end{gathered} 
\label{ec:solution}
\end{equation} 

Here, 
\begin{equation} 
\begin{gathered} 
\lambda^{2} = \frac{\sigma_{n} + \sigma_{p}}{\sigma_{n}\sigma_{p}} 
\frac{n_{0}(T^{*})p_{0}(T^{*})}{n_{0}(T^{*}) + p_{0}(T^{*})} 
\frac{e^{2}}{\tau T^{*}} ,\\ 
\beta^{2} = 
\frac{\sigma_{n}p_{0}(T^{*}) - \sigma_{p}n_{0}(T^{*})}%
{\sigma_{n}\sigma_{p} 
\left[ 
n_{0}(T^{*}) + p_{0}(T^{*}) 
\right]} 
\frac{e^{2}n_{0}(T^{*})}{\tau T^{*}} . 
\end{gathered} 
\end{equation} 

Constants $C_{1,2,3,4}'$ should be determined from boundary conditions
(\ref{ec:front1}) and (\ref{ec:front2}). 

After some mathematical manipulation, we arrive to general expressions
for the total current in bipolar semiconductors: 
 \begin{equation*} 
j_0\left( \frac{2}{\sigma _n^s}+\frac {L}{\sigma _m} + 
\frac{2a}{\sigma_n +\sigma_p} 
\left[1 + \frac{\sigma_p/\sigma_n}{\lambda a \coth\left( \lambda a \right) + \lambda^2\tau a S }\right] \right) = 
2a\left[ \frac{E_1}{\lambda a \coth\left( \lambda a \right) + \lambda^2\tau a S} - E\right],
\end{equation*}
where 
\begin{equation} 
\begin{gathered} 
E_1 =\frac{\sigma_p\left(\alpha_p-\alpha_n\right)}{ 
\sigma_p+\sigma_n} \frac{dT}{dx} , \\ 
E = \frac{\alpha_{n}\sigma_{n} + 
\alpha_{p}\sigma_{p}}{\sigma_{n} + \sigma_{p}} \frac{dT}{dx} 
\end{gathered} 
\end{equation} 

Terms $2/\sigma _n^s$ and $L/\sigma _m$ determine the contact electric 
resistance and the resistance of metal section, respectively. 

Let us compare now this expression with the general Ohm's law 
\begin{equation} 
j_{0} {\cal R} = \cal E , 
\end{equation} 
where ${\cal R}$ is the total circuit resistance per cross section unit, 
$\cal E$ is e.m.f. 

We can state that, 
\begin{equation} 
R_s = \frac{2a}{\sigma_n +\sigma_p} 
\left(1+\frac{\sigma_p/\sigma_n}{\lambda a \coth\left(\lambda a\right) + \lambda^2\tau a S} \right) 
\label{ec:res} 
\end{equation} 
is the bipolar semiconductor resistance, and 
\begin{equation} 
{\cal E} = 2a\left[ \frac{E_1}{\lambda a \coth\left( \lambda a \right) + \lambda^2\tau a S }- E \right] 
\end{equation} 
is the thermo-e.m.f. 

The electron and hole Fermi quasi-levels have the following general
form: 
\begin{equation} 
\begin{gathered}
\tilde{\varphi}_{n} 
= -a
\left[
\left(
\frac{j_{0}}{\sigma_{n}+\sigma_{p}} + E 
\right) 
\frac{x}{a} 
- \frac{j_{0}L}{2\sigma_{m}a} + 
\frac{j_{0}\frac{\sigma_{p}}{\sigma_{n}(\sigma_{n}+\sigma_{p})} - 
E_{1}}{a\lambda\coth \lambda a + \lambda^2 a\tau S}
\frac{\sinh\lambda x}{\sinh\lambda a}
\right] \label{ec:fql} \\ 
\tilde{\varphi}_{p}
=
-a
\left[
\left(\frac{j_{0}}{\sigma_{n}+\sigma_{p}} + E \right) 
\frac{x}{a} - \frac{j_{0}L}{2\sigma_{m}a} -
\frac{j_{0}\frac{1}{\sigma_{n}+\sigma_{p}} - 
\frac{\sigma_{n}}{\sigma_{p}}E_{1}}{a\lambda\coth 
\lambda a + \lambda^2 a\tau S}
\frac{\sinh\lambda x}{\sinh\lambda a}
\right] 
\end{gathered}
\end{equation} 
which shows that they are really different and this confirms our initial 
assumptions. 

From expressions similar to Eqs.~(\ref{ec:dn}, \ref{ec:dp}) and 
(\ref{ec:solution}), it is easy to obtain the concentrations of 
nonequilibrium carriers,
\begin{equation} 
\delta n(x) = \delta p(x) = \frac{ea}{T^{*}}\frac{n_{0}(T^{*})p_{0}(T^{*})}{n_{0}(T^{*}) + p_{0}(T^{*})}\frac{\sigma_{n} + \sigma_{p}}{\sigma_{n}}
\frac{j_{0}\frac{1}{\sigma_{n}+\sigma_{p}} - \frac{\sigma_{n}}{\sigma_{p}}
E_{1}}{a\lambda\coth \lambda a + \lambda^2 a\tau S}
\frac{\sinh\lambda x}{\sinh\lambda a}
\end{equation} 

It is important to emphasize that in the proposed approach to 
thermoelectric transport, thermo-e.m.f. depends not only on electron
and hole thermoelectric powers and electric conductivities, but also on
the carriers' life time and surface recombination rate. If bulk and/or
surface recombination processes are intensive enough ($S\gg S_0 =
\frac{n_0+p_0}{\sigma_{n}+\sigma_{p}}\frac{T\sigma_{p}^2}{e^2 a
n_{0}p_{0}}$ and/or $\tau\ll\tau_{0}= \frac{e^2 a
n_{0}}{T}\frac{\sigma_{n}+\sigma_{p}}{\sigma_{n}\sigma_{p}}$) then the
thermo-e.m.f. is given by
\begin{equation} 
{\cal E} 
=
-2a 
\frac{\alpha_{n}\sigma_{n} + \alpha_{p}\sigma_{p}}{\sigma_{n}+\sigma_{p}}
\frac{d T}{dx} 
\end{equation} 
i.e. we obtain the well-known result\cite{Anselm} for thermo-e.m.f.

On the contrary, when surface and bulk recombinations are quite weak 
($S\ll S_{0}$, $\tau\gg\tau_{0}$) then we arrive to another 
result: 
\begin{equation} 
{\cal E}=-2a\alpha_{n}\frac{dT}{dx} 
. 
\label{ec:el} 
\end{equation} 

It is important to note that result (\ref{ec:el}) concerns both
$n$-type and $p$-type semiconductors.

Only electrons take part in producing a thermo-e.m.f. in the absence of
recombination. This can be this way, only because $j_{p}=0$ under $R=0$,
$R_{s}=0$ [see Eqs.~(\ref{ec:pcurrent},\ref{ec:front1})], and the total
current $j_{0}$ coincides with the electron current $j_{n}$. Thus, the
thermo-e.m.f. can change sign in $p$-type semiconductors when both surface
and bulk recombination rates decrease.

Let us note that it is possible to rewrite the condition of weak
recombination ($\tau\gg\tau_{0}$) and weak surface recombination ($S\ll
S_{0}$) in the form 
\begin{equation} 
\begin{gathered} 
a\ll\lambda^{-1},\\ 
S\ll\frac{lv_{T}}{a} 
\end{gathered} 
\label{ec:aS} 
\end{equation} 
respectively, where $l$ is momentum free length, $v_{T}$ is the thermal
mean velocity. 

As it was shown in Ref.~\onlinecite{Borodovski}, surface
recombination velocity in epitaxial layers $n-Cd_{x}H_{1-x}Te$ is
$S\sim10^3$~cm/s at room temperature. As $v_{T}\sim 10^7$~cm/s, $l\sim
10^{-5}$--$10^{-6}$ cm, surface recombination becomes non-effective
if $2a\ll 10^{-2}$ cm [see Eq.~(\ref{ec:aS})] (from the point of view of
non-equilibrium carriers appearance). If the condition
$2a\ll\lambda^{-1}$ takes place, at the same time, the non-equilibrium
carriers will play the main role in formation of thermo-e.m.f. 

We attract attention upon the fact that bipolar semiconductor resistance
and thermo-e.m.f. depend not only on electron and hole electrical
conductivities, as it is usually considered, but on recombination rates
as well.

The semiconductor resistance takes the usual form: 
\begin{equation} 
R_{s} = \frac{2a}{\sigma_{n}+\sigma_{p}} 
,
\end{equation} 
only in the case of strong bulk and/or surface recombinations. 
In general, it is impossible to calculate the semiconductor resistance
independently of full current and thermo-e.m.f. 

It should be mentioned that, within the framework of the model
considered here, the semiconductor resistance takes the same form as in
Eq.~(\ref{ec:res}) in the absence of temperature gradient but in the 
presence of an external voltage.\cite{FTP} 

\begin{figure}[!ht]
\begin{center}
\includegraphics[width=5cm, height=8cm, angle=-90]{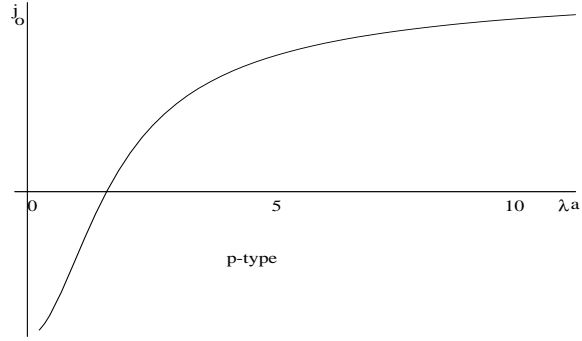}
\end{center}
\caption{%
$j_{0}$ as function of $\lambda a$ when $S=0$ in arbitrary units.}
\label{f:fig9}
\end{figure}

\begin{figure}[!ht]
\centering
\includegraphics[width=5cm, height=8cm, angle=-90]{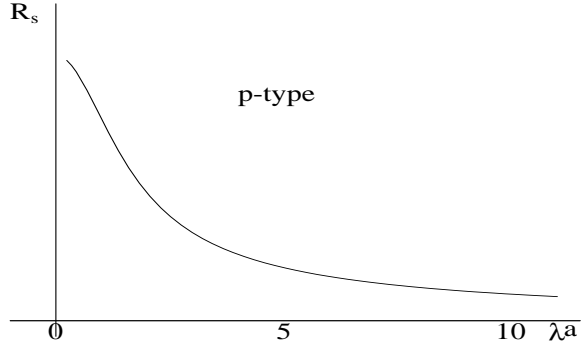}
\caption{%
$R_{s}$ as function of $\lambda a$ when $S=0$ in arbitrary units.}
\label{f:fig10}
\end{figure}

\begin{figure}[!ht]
\centering
\includegraphics[width=5cm, height=8cm, angle=-90]{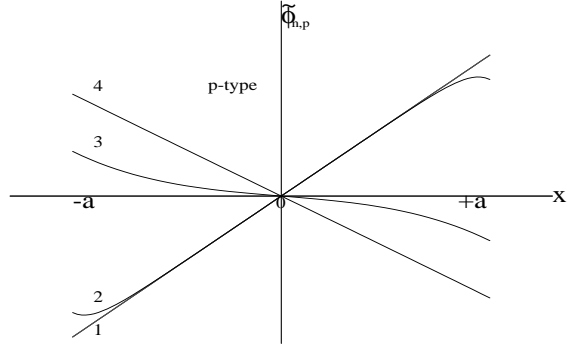}
\caption{%
Fermi quasi-levels for holes $\tilde\varphi_{p}(x)$ (1) and
electrons $\tilde\varphi_{n}(x)$ with small (2), intermediate (3), and big
(4) life time when $S=0$ in arbitrary units.}
\label{f:fig11}
\end{figure}

Figs.~\ref{f:fig9} 
and \ref{f:fig10} 
show the bulk recombination ($\lambda a \sim 1/\sqrt{\tau}$) dependence 
of total current $j_{0}$ and semiconductor resistance $R_{s}$ 
respectively. The dependence of Fermi quasi-levels on the surface ($S$) 
and bulk recombination is illustrated in Fig.~\ref{f:fig11}. As seen from 
Figs.~\ref{f:fig9}-\ref{f:fig10}, the total current and the semiconductor 
resistance depend essentially on the electron-hole recombination rate. It 
is important to note that for a p-type semiconductor the total current 
can change direction in dependence of the recombination rate.

Let us pay attention that the Fermi quasi-level of electrons becomes a
non-monotonous function of coordinates [see curve (2) in
Fig.~\ref{f:fig11}] 
under small bulk and surface recombination rates. As
far as we know, it is the only case when Fermi quasi-level depends on
coordinate nonmonotonically. 

\section{Conclusion} The area of application of the stated theory is wider
and comprehends more complex problems than those solved in the present
work. For example when current flows in a closed circuit formed by a
current source, $p$-type semiconductor and connecting metal wires, it is
necessary to take into account the effects described in the present
work.\cite{FTP} The above stated theory shows that essential increase of
temperature in $n$-type semiconductor with large life times and small
surface recombination rates does not lead to noticeable changes of
thermo-e.m.f. That happens because holes do not contribute to the
formation of Seebeck effect, even if its concentration is large enough.
This fact can be of essential importance to increase the figure of merit
of thermoelectric devices (thermo-generators, thermo-refrigerators).

In general, electron and hole temperatures do not coincide and are not
linear functions of coordinates,\cite{Bochkov} so their temperature
distributions should be determined by solving the set of energy balance
equations for subsystems of electrons, holes, and phonons with the
appropriate boundary conditions.

Let us note that the ideas stated in works
[\onlinecite{Lyubimov,Gurevich}] and developed in the present work have
found an ulterior development in Ref.~[\onlinecite{Konin,Raguotis}]. At
present, experimental works\cite{Vackova,KK,VackJT} have appeared based in
the model presented by us.  At present, these first results are the
initial based on this theory and allow us to think in optimizing the
parameters of thermoelectric devices.

\section*{ACKNOWLEDGMENTS} 
The authors would like to thank CONACyT, M\'{e}xico, for partial support.

\end{document}